\documentclass[twocolumn,showpacs,preprintnumbers,amsmath,amssymb,superscriptaddress]{revtex4-1}
\usepackage{amsmath}
\usepackage{graphicx}
\usepackage{amsfonts}
\usepackage{amssymb}
\usepackage{bm}
\usepackage{color}%
\usepackage{ulem}

\begin{document}

\title{\boldmath{$\beta$NMR study of Isolated $^8$Li$^+$ in the Enhanced Paramagnet Platinum}}

\author{Oren~Ofer}
 \email{oren@triumf.ca}
\affiliation{TRIUMF, 4004 Wesbrook Mall, Vancouver, BC, V6T 2A3 Canada}
\author{K.H.~Chow}
\affiliation{Department of Physics, University of Alberta, Edmonton, AB, T6G 2E1 Canada}
\author{I. Fan}
\altaffiliation{Department of Physics, National Tsing Hua University, Hsinchu, Taiwan, R.O.C. 30013}
\affiliation{Department of Physics, University of Alberta, Edmonton, AB, T6G 2E1 Canada}
\author{M. Egilmez}
\altaffiliation{Department of Materials Science and Metallurgy, University of Cambridge, Cambridge,
United Kingdom, CB2 3QZ}
\affiliation{Department of Physics, University of Alberta, Edmonton, AB, T6G 2E1 Canada}
\author{T.J. Parolin}
\affiliation{Department of Chemistry,  University of British Columbia, Vancouver, BC, V6T 1Z1 Canada}
\author{M.D. Hossain}
\affiliation{Department of Physics and Astronomy, University of British Columbia, Vancouver, BC V6T 1Z1 Canada}
\author{J. Jung}
\affiliation{Department of Physics, University of Alberta, Edmonton, AB, T6G 2G7 Canada}
\author{Z. Salman}
\affiliation{Paul Scherrer Institute, Laboratory for Muon Spin Spectroscopy, 5232 Villigen PSI, Switzerland}
\author{R.F. Kiefl}
\affiliation{TRIUMF, 4004 Wesbrook Mall, Vancouver, BC, V6T 2A3 Canada}
\affiliation{Department of Physics and Astronomy, University of British Columbia, Vancouver, BC V6T 1Z1 Canada}
\affiliation{Canadian Institute for Advanced Research, CIfAR}
\author{C.D.P. Levy}
\affiliation{TRIUMF, 4004 Wesbrook Mall, Vancouver, BC, V6T 2A3 Canada}
\author{G.D. Morris}
\affiliation{TRIUMF, 4004 Wesbrook Mall, Vancouver, BC, V6T 2A3 Canada}
\author{M.R. Pearson}
\affiliation{TRIUMF, 4004 Wesbrook Mall, Vancouver, BC, V6T 2A3 Canada}
\author{H. Saadaoui}
\affiliation{Paul Scherrer Institute, Laboratory for Muon Spin Spectroscopy, 5232 Villigen PSI, Switzerland}
\author{Q. Song}
\affiliation{Department of Physics and Astronomy, University of British Columbia, Vancouver, BC V6T 1Z1 Canada}
\author{D. Wang}
\affiliation{Department of Physics and Astronomy, University of British Columbia, Vancouver, BC V6T 1Z1 Canada}
\author{W.A.~MacFarlane}
\affiliation{Department of Chemistry,  University of British Columbia, Vancouver, BC, V6T 1Z1 Canada}
\date{\today}

\begin{abstract}
We report $\beta$ detected nuclear  magnetic resonance ($\beta$NMR) measurements of $^8$Li$^+$
implanted into high purity Pt. The frequency of the $^8$Li $\beta$NMR resonance and the
spin-lattice relaxation rates $1/T_1$ were measured at temperatures ranging from 3 to 300~K.
Remarkably, both the spin-lattice relaxation rate and the Knight shift $K$ depend linearly on temperature $T$ although
 the bulk susceptibility does not. $K$ is found to scale with the Curie-Weiss dependence of the
Pt susceptibility extrapolated to low temperatures. This is attributed to a defect response of the enhanced paramagnetism of Pt, i.e.
the presence of the interstitial Li$^+$ locally relieves the tendency for the Curie-Weiss susceptibility to saturate at low $T$.
We propose that the low temperature saturation in $\chi$ of Pt may be related to an interband coupling between the $s$ and $d$
bands that is disrupted locally by the presence of the Li$^+$.  
\end{abstract}


\maketitle

\section{Introduction}

In metals, enhanced paramagnetism is often
observed as a large temperature-dependent paramagnetic susceptibility.
Various compounds \cite{titberil,SrRhO} as well as elemental metals \cite{cesium,pd}
have been reported to exhibit such intriguing behavior,
motivating significant theoretical efforts
towards developing understanding of these phenomena, leading
to theoretical models such as self-consistent
renormalization (SCR) \cite{scrFM,scrAFM} and other theories \cite{dosModels,spinFluct}.
In fact, the SCR models have recently been applied to antiferromagnetic doped cuprates \cite{adapt}
and itinerant magnetic systems \cite{adapt2}. The underlying context of these theories is the proximity
to a quantum critical point (QCP).
It has been recently predicted that dilute impurities, such as Co, in a nearly ferromagnetic host,
such as Platinum (Pt), could serve as a test case for the behavior of magnetic 
droplets in the vicinity of a QCP \cite{loh}.
\begin{figure}[tbh]
\begin{center}
\includegraphics[height=\columnwidth,angle=-90]{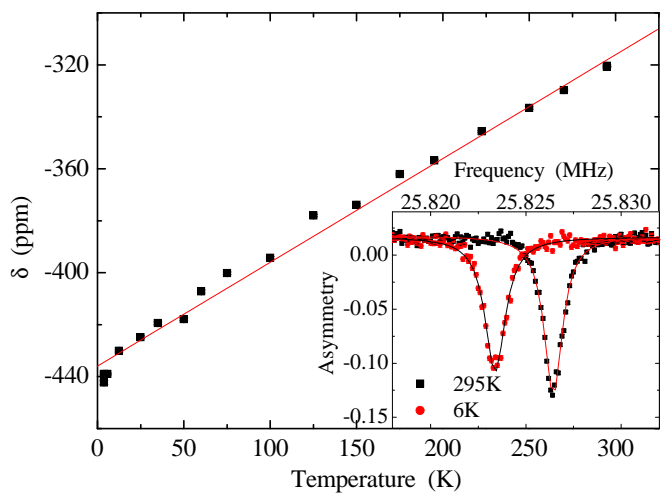}
\caption{(Color online)
The temperature dependence of the
frequency shift
$\delta$  of $^8$Li in Pt at an applied field of 4.1 T.
The inset shows the resonance spectra of
$^8$Li in Pt at 6~K (red) and 295~K (black) with fits described in the text.}
\label{fig:delta}
\end{center}
\end{figure}

\begin{figure}[tb]
\includegraphics[width=\columnwidth]{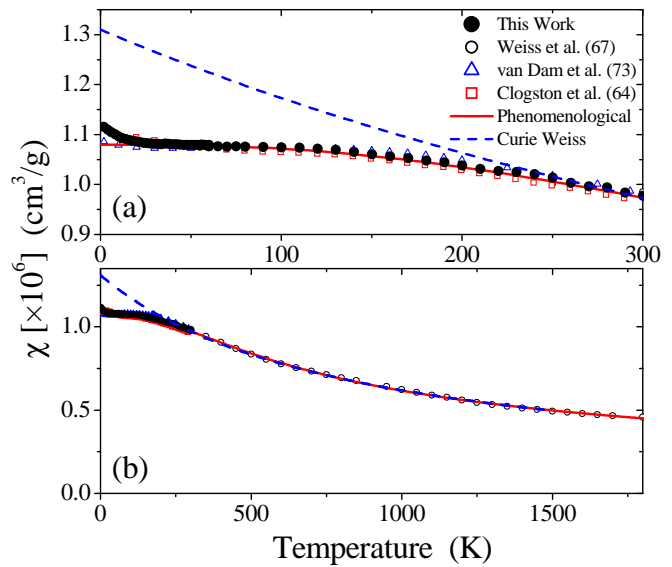}
\caption{(Color online) The measured SQUID magnetic susceptibility versus temperature
for the Pt foil sample for (a) temperatures below 300 K and (b) to high temperatures
(filled black circles) in comparison with literature values \cite{chi-refs}.
The small Curie-like upturn below 20 K is likely due to a low level of residual
magnetic impurities. The curves are fits to the high temperature CW law (blue dashed) and
to a form that saturates at low temperature (red solid)\cite{chi-fit}.
}
\label{fig:squid}
\end{figure}

Pt is an enhanced paramagnet that has magnetic properties unlike noble metals such as
Au, Ag, or Cu.  Instead, it is closely related to Pd.
Pt has a strong electron exchange interaction that results in an
enhanced paramagnetic susceptibility $\chi$ that is temperature dependent, exhibiting Curie-Weiss (CW) behaviour
above $\approx 300$ K and saturation or possibly a weak maximum below this [see Fig.~\ref{fig:squid}(b)],
features shared with Pd and other nearly ferromagnetic metals.
The origin of the CW $\chi$ is still controversial. One line of theoretical explanation holds that it is purely a band structure
effect related to the strongly energy dependent density of electronic states at the Fermi level, $\rho(E_F)$ \cite{bands}.
Another explanation relies on the importance of spin fluctuations \cite{paramag,moriya-book}, while a third gives priority to the
electron-phonon interaction \cite{kim88}. The low temperature saturation (or peak) in $\chi$ has also been debated extensively
and may be explained in each of the above pictures.
In the context of nearly ferromagnetic metals, Pt is less magnetic than Pd, with
susceptibility nearly an order of magnitude smaller.
Also, unlike Pd, relativistic effects are important in its electronic structure \cite{relative}.

Here we report $^{8}$Li beta-detected  nuclear magnetic resonance ($\beta$NMR) and relaxation
 rate measurements in Pt.  The $\beta$NMR technique gives us the unique opportunity to explore several aspects of the implantation
of a spin polarized radioactive $^8$Li$^+$ ion in Pt.
One question is whether the $^8$Li will locally perturb the Pt host
or whether it acts as a passive probe revealing the intrinsic properties of the host.
Second, the measurements presented here can provide a useful reference for future $\beta$NMR experiments
on heterostructures with Pt overlayers, finite size effects in low-dimensional Pt (including the
recently reported magnetism in Pt nanoparticles \cite{Ptnano}),
as well as depth-controlled experiments on samples that use Pt as a substrate,
or a component layer in a multilayer structure.
A detailed understanding of the interaction of alkali metals with Pt may 
also help to clarify their role in chemical catalysis \cite{norcross99}.
Our main findings are that both the $^{8}$Li spin-lattice
relaxation rate and (surprisingly) Knight shift vary linearly
from room temperature down to 3 K, although the bulk susceptibility does not.
We discuss this discrepancy and compare the $\beta$NMR results with another implanted local probe,
the positive muon, by making similar muon spin rotation ($\mu$SR) measurements of the Knight shift on the same sample.
The $\mu$SR results are less accurate because of the much shorter muon lifetime, and they are compatible with either 
the bulk $\chi$ or the $^8$Li shift.

\section{The Experiment}

The Pt sample is a high purity ($99.999\%$) foil (0.001" thick) weighing $51.3471$~mg \cite{espi},
folded into a $5\times5$mm square.
The foil was initially annealed \cite{ohtsubo} at 800 $^\circ$C for 5 hours in oxygen
and, prior to the $\beta$NMR experiments,
was thoroughly cleaned with methanol (Fisher Scientific).
The $^{8}$Li $\beta$NMR was carried out on the polarized low energy beamline at
TRIUMF's ISAC facility \cite{BNMR:zaher,BNMR:kim}  where
highly polarized ($>70\%$) $^8$Li$^+$ probe nuclei (nuclear spin $I=2$,
lifetime $\tau$=1.21 s, gyromagnetic ratio $^8\gamma=6.3015$~MHz/T) are implanted into
 the Pt foil.
 The  experiments were carried out at an implantation energy of $28$ keV,
 corresponding to a range of $\approx 61$ nm and a straggle of $\approx 33$ nm.
 The asymmetry $\mathcal{A}(t)$ of the $\beta$-decay electron emission, proportional to the $^{8}$Li nuclear
spin polarization, is recorded.
For the resonance measurements, a continuous beam of $\approx10^7$ ions/s is
focused into a beam spot of $\approx3$ mm in diameter.
 A small pulsed radio-frequency magnetic field ($H_1\sim100\mu$T) is applied transverse to the spin polarization.
The central frequency $\nu$ of the rf pulses is randomly swept through an interval, and resonance is achieved when the
Larmor condition, i.e. rf frequency equal to $^8\gamma B_{int}$, is satisfied,
 manifesting itself as a loss of the time-averaged $A(t)$ at that frequency.
The spin-lattice relaxation rate $T_1^{-1}$ is measured in a pulsed experimental mode using a
4~s beam-on pulse, followed by a 12~s counting period with beam off.
This cycle is repeated on the order of 100 times to accumulate statistics for each measurement.
Here, the rf field is off at all times, and the time dependence of $\mathcal{A}(t)$ is monitored both during and after
the pulse.

The $\mu$SR data were collected at the M15 surface muon beamline at TRIUMF using a low background insert
in the HELIOS spectrometer with an applied transverse field of $H_0=1.5~$T. The precession
signal was weakly relaxing and
was fit to a single exponentially damped oscillation with a small damping rate ($\approx 0.095$~$\mu$s$^{-1}$) that is
largely due to the random dipolar fields of the host $^{195}$Pt nuclei.
This rate was approximately independent of temperature (within $\approx 10$\%).
Reference spectra in pure Ag were collected simultaneously and used as the reference for the shift,
accounting for the known Knight shift of $\mu^+$ in Ag \cite{schenckbook}.

Bulk susceptibility measurements were performed using a Quantum Design Magnetic Property Measurement
System superconducting quantum interference device (SQUID) under the same field as the resonance measurements
($H_0=4.1$~T).

\section{Results}

The inset of Fig. \ref{fig:delta} shows typical resonance spectra of $^{8}$Li$^+$ in Pt at $295$~K and $6$~K.
In contrast to previous $\beta$NMR studies on other FCC metals (Ag \cite{AndrewAuAg,GerraldAg}, Au \cite{AndrewAuAg,TerryAu},
Cu \cite{zaherCu,LCR-Cu}  and Pd \cite{TerryPd}), we find a single narrow resonance at all temperatures, with no evidence of multiple sites for
the $^{8}$Li$^+$ below 300~K.  The observation of a single narrow resonance close to the Larmor frequency with no
quadrupolar splitting establishes that $^8$Li is located at a cubic site, which, in the FCC lattice could be the substitutional,
the octahedral interstitial, or the tetrahedral interstitital site.
By analogy with the other FCC metals, it is very likely that below 300~K, $^8$Li$^+$
occupies the octahedral interstitial site. We expect that at higher temperatures, with sufficient thermal energy, the $^8$Li will
find a nearby vacancy and become substitutional, i.e. the site transition observed in other FCC metals 
occurs above 300~K. This is similar to Pd and consistent with the higher formation energy for a vacancy
(and lower vacancy mobility) in Pt compared to the other metals we have studied \cite{ehrhart}.  

The resonance at all temperatures is well-described by a Lorentzian,
\begin{equation}
\mathcal{A}(\nu)=\frac{\mathcal{A}_0}{2\pi}\frac{\Delta \nu } {(\nu-\nu_{Pt})^2+\Delta \nu^2}\label{lorentz}
\end{equation}
where $\mathcal{A}_0$ is the amplitude, $\nu$ and $\nu_{Pt}$ are the applied rf and the Pt resonance frequencies, respectively.
$\Delta \nu$ is the full-width at half-maximum (FWHM).
The resonance linewidth (not shown), $\Delta \nu=1.2(1)$~kHz is temperature independent,
consistent with a single site for $^8$Li in this temperature range. The main panel of Fig.~\ref{fig:delta} shows
the temperature dependence of the Pt frequency shift, $\delta=(\nu_\text{Pt}-\nu_\text{MgO})/\nu_\text{MgO}$
where $\nu_\text{MgO}=25.834738(3)$~MHz is the
resonance frequency of $^8$Li in a single crystal MgO reference sample.
The reference measurement was carried out both
before and after the Pt measurements to ensure the stability of the external field $H_0$,
and to determine the exact field at the sample position.

\begin{figure}[tbp]
\begin{center}
\includegraphics[width=\columnwidth]{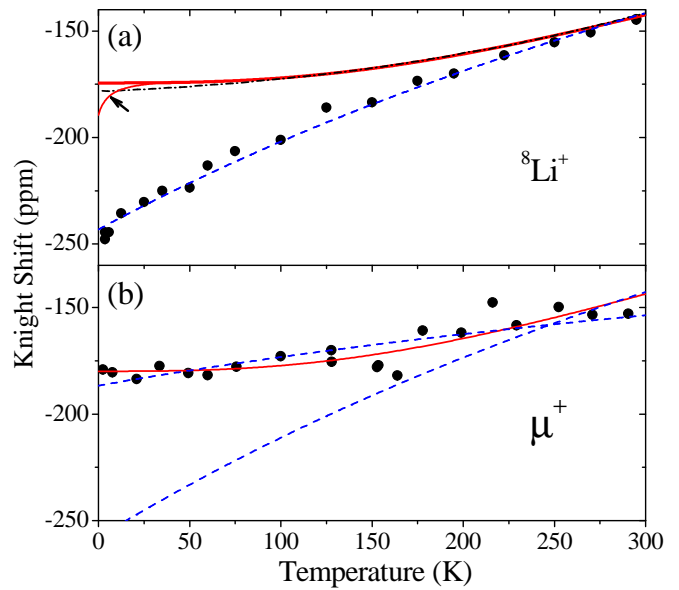}
\caption{(Color online) The temperature dependence of the (a) $^8$Li and (b) muon Knight shifts in the same Pt sample.
(a) A linear fit to $\chi(T)$ above 250 K [Eq. (\ref{jacc})] yields the heavy red curve which deviates from the data.
Using the same fit, but substituting $\chi_{CW}(T)$, reproduces the data well over the whole range (blue dashed curve).
A similar scaling with the $^{195}$Pt Knight shift \cite{shaham} yields the dot-dash black curve.
The arrow indicates the measured Curie-like contribution to $\chi$, which, if due to dilute magnetic impurities,
will not contribute to the shift. (b) In contrast, $K(T)$ for $\mu^+$ fits the bulk $\chi_p(T)$ (red curve).
Using the same fit parameters with $\chi_{CW}$ yields the steep dashed blue curve, while fitting instead over
the full temperature range provides nearly as good a fit as $\chi_p(T)$ (shallow blue dashed curve).
}
\label{fig:K}
\end{center}
\end{figure}

Figure \ref{fig:delta} shows that $\delta$ is linear in $T$ and furthermore,
it is negative and monotonically decreasing in magnitude with increasing $T$.
This is similar to Pd \cite{TerryPd} but different from other noble metals where a positive
$T$-independent shift is observed \cite{zaherCu,GerraldAg,TerryAu}.
The negative shift is attributed to the hyperfine coupling due to hybridization of the (nearly) vacant Li $2s$ orbital with the
host $d$ band, i.e. ``$s$-$d$ hybridization" \cite{akai-rev}.
In contrast, when the $d$ band is filled, as in Ag, for example, where it lies significantly below the Fermi level,
the hyperfine coupling to the implanted between $^8$Li and the partly filled $s$ band is positive.

The NMR shift provides a local measure of the spin susceptibility, and to analyze it further, we require an
independent measurement of the bulk susceptibility $\chi(T)$. To this end, we carried out a SQUID measurement of $\chi$
on the same sample.
The results, after subtracting a temperature independent sample holder contribution ($1.1\times 10^{-7}$ cm$^3$/g),
are shown in comparison to conventional values from the literature \cite{chi-refs} in Fig. \ref{fig:squid}.
The measured $\chi$ is consistent with the standard temperature dependence, exhibiting a saturation of the high temperature
CW law below room temperature.
$\chi$ also shows a small Curie-like upturn at low temperature (below 20 K)
that is attributed to a low concentration of
residual magnetic impurities. To facilitate comparison of our results with the bulk susceptibility, we fit $\chi(T)$ to
a CW law (at high temperature) and to a phenomenological form $\chi_p(T)$ that accounts for the
low temperature saturation \cite{chi-fit}, see curves in Fig. \ref{fig:squid}.

We use the measured $\chi(T)$ (in fact, $\chi_p$) to obtain the Knight shift by correcting for demagnetization.
Assuming the Pt foil behaves as a thin slab \cite{XuDemag}, $K(T)=\delta(T)+(8\pi/3)\rho/M_r\cdot\chi(T)$, where
 $\rho$ and $M_r$ are the density and the molar mass of Pt. $K(T)$ thus calculated is about 200 ppm
smaller in magnitude than $\delta$ and is shown in Fig.~\ref{fig:K}(a).
Since $\chi$ is almost $T$-independent below 300~K (Fig.~\ref{fig:squid}), the correction to $\delta$ is also nearly constant, and
the temperature dependence of $K$ shows only a small difference, exhibiting a slight curvature.

Knowledge of $K(T)$ enables determination of the local electronic structure around
the implanted Li through the hybridization of its local ($2s$) electron with the Pt conduction band(s).
This hybridization leads to a hyperfine coupling $A$.
In the NMR of transition metals, $K$ is usually decomposed into three (assumed independent) contributions
 \begin{equation}
 K=K_{\rm s}+K_{\rm d}(T)+K_{\rm orb}.
\label{shiftcomps}
 \end{equation}
The first term is from the broad $s$ band, the second from the narrow $d$ band, and the third, from orbital screening (the chemical shift),
see Ref. \onlinecite{clogston}, for example. Only the second term is assumed to have any temperature
dependence. We note that hybridization mixes the character of the conduction bands in Pt. While three bands cross the Fermi level,
only two make a substantial contribution to the density of states. One of these (the $\Gamma6$ band) is broad while the other
($XW5$) is substantially narrower, contributing about 80\% of the density of states at $E_F$ \cite{Pt-bands,dkc78}. 
We identify these as the $s$ and $d$ bands, respectively, in our simplified discussion.

We expect a similar decomposition of the shift applies to the $^8$Li impurity hyperfine-coupled to the host conduction bands, but
the relative magnitudes of each term will be substantially different than for the host nuclear spins. In particular, 
with a small nuclear charge and only a few electrons, the orbital shift for $^8$Li is generally small (less than 10 ppm).
In FCC metals with filled $d$ bands, $K_d=0$ and $K_s$ for $^8$Li is on the order of +100 ppm \cite{TerryAu}, so
we may expect such a shift from the $s$ band in Pt.
However, with the present data, rather than using an uncertain decomposition of the temperature independent terms, we adopt the
following simplified relation
\begin{equation}
K(T) = A \chi(T) + \kappa,
\label{jacc}
\end{equation}
where $K$ is the corrected Knight shift and $\chi$ is the measured total susceptibility (without the impurity contribution, see below).
Any difference in the hyperfine coupling to the different bands is then accommodated by the constant $\kappa$.
In order to extract the hyperfine coupling, we fit the observed $K(T)$ to Eq. (\ref{jacc}). However, as is evident in the
raw data (Fig. \ref{fig:delta}), the shift does not saturate to a constant value below 300~K, and the fit using $\chi_p$
(not shown) is poor. Above 250~K, where $\chi_p$ starts to show some $T$ dependence, we can obtain
a reasonable fit and extract the hyperfine coupling, but the curve [red curve in Fig.~\ref{fig:K}(a)] deviates from the data at lower temperature.
Using this same coupling [$A$ and $\kappa$ from the fit to Eq. (\ref{jacc})] but $\chi_{CW}$ instead,
yields the blue dashed curve that fits the data well down to base temperature.
From the fit, the hyperfine coupling for $^8$Li is $A = -8.6(0.9)$ kG/$\mu_B$.
With the assumption that the temperature dependence of $\chi$ is entirely from the $d$ band,
$A$ is the coupling of the $^8$Li to the $d$ band.
Note the value is not divided by the number of nearest neighbours to give a
per neighbour atomic value, rather it is the net value for the $^8$Li at this site. $A$ is thus substantially smaller
in magnitude than the couplings to the $s$ band in the noble metals \cite{TerryAu}, but is similar to Pd, where 
$s$-$d$ coupling is also important \cite{TerryThesis}.

In contrast, the Knight shift of the positive muon, though less accurate due to the short muon lifetime,
fits well to the bulk $\chi$ [red curve in Fig.~\ref{fig:K}(b)], with a hyperfine coupling $A = -9.8(3.2)$ kG/$\mu_B$, roughly
consistent with a previous report \cite{gygax1} and similar to the $^8$Li.
Using this fit and $\chi_{CW}$ yields
the steep dashed blue line that clearly deviates from the data.
It should be noted, however, that due to the significant scatter,
one can also fit the muon shift to $\chi_{CW}$ nearly as well as to the saturating $\chi_p$,
see the shallow blue dashed curve in Fig.~\ref{fig:K}, with a substantially smaller hyperfine coupling $A = -2.8(0.5)$ kG/$\mu_B$.
The hyperfine couplings for both the charged implanted probes are thus small, negative, and (at least for the saturating fit of the muon shift)
similar in magnitude. However, the $^8$Li, at least, clearly shows a distinct $T$
dependence to the shift that is not consistent with the bulk $\chi(T)$.

If the small Curie-like contribution in $\chi$ is due to dilute magnetic impurities it should not
contribute to the shift. Instead, it should cause an additional inhomogeneous magnetic broadening scaling with
that term in $\chi$ \cite{WW74}. Fig.~\ref{fig:K} shows there is no significant contribution to the shift.
We can roughly estimate the broadening effect by noting that this term in $\chi$ is consistent with
the Curie law for $\approx10$ ppm Fe impurities with moment  $6.5\mu_B$.
This yields an estimated magnetic dipolar line broadening of $\sim0.3$ kHz at 6 K \cite{WW74},
a small fraction of the observed $^8$Li resonance width, consistent with the observed temperature independence of the linewidth.

\begin{figure}[tbp]
\begin{center}
\includegraphics[height=\columnwidth,angle=-90]{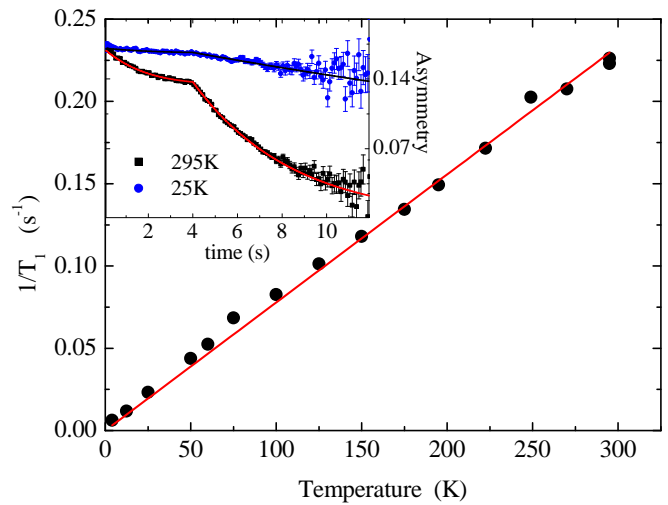}
\caption{(Color online) The temperature dependence of the $T_1^{-1}$ of $^8$Li in Pt. The fit of the
data to a straight line through the origin yields $T_1^{-1}=7.78(9)\times10^{-4}$s$^{-1}$K$^{-1}T$.
 The inset shows the $^8$Li time-dependent asymmetry at 25 K (red) and 295~K (black); the lines
  indicate the fit to Eq.(\ref{equ:t1}).}
\label{fig:t1}
\end{center}
\end{figure}

We now turn to the $^8$Li spin-lattice relaxation that is determined by electronic spin fluctuations.
In the inset of Fig.~\ref{fig:t1} we plot representative data at two temperatures,
25 K and 295~K, in 4.1 T applied field.
Clearly, the relaxation rate, $T_1^{-1}$, decreases with decreasing temperatures.
Recall that the data are collected during and after a 4 second $^8$Li$^+$ beam pulse.
Assuming a single exponential spin lattice relaxation, the observed polarization
is fit to the bipartite relaxation function, during $\mathcal{A}_d(t)$ and after $\mathcal{A}_a(t)$ the pulse \cite{masrurRelax},
\begin{align}
\mathcal{A}_d(t) & =\mathcal{A}_0\frac{\tau'}{\tau}\frac{1-\exp(-t/\tau')}{1-\exp(-t/\tau)}\nonumber\\
\mathcal{A}_a(t) & =\mathcal{A}_d(\Delta)\exp(-(t-\Delta)/T_1)
\label{equ:t1}
\end{align}
where $(\tau')^{-1}=\tau^{-1}+T_1^{-1}$ and $\Delta=4$ s. The relaxation rates $T_1^{-1}$
extracted from the fits are shown in the main plot of Fig.~\ref{fig:t1}, indicating that the
reduction in $T_1^{-1}$ with $T$ is monotonic and linear.
The slope, $1/T_1T$, is found to be comparable to that in other noble metals \cite{TerryPd,zaherCu,GerraldAg,TerryAu}.
Such a temperature dependence is characteristic of relaxation due to spin flip scattering between the nucleus and the
conduction electrons of a metal, i.e. the Korringa mechanism. Note that analogous measurements for the muon
are not possible as the Korringa relaxation time is much longer than the 2.2 $\mu$s muon lifetime.

\section{Discussion}

The absence of saturation in $K(T)$ for $^8$Li is puzzling. This feature is thought to be intrinsic to the
electronic susceptibility of Pt, so it should be reflected in the shift. This appears to be the case for the 
host $^{195}$Pt NMR shift \cite{clogston,shaham}, though detailed low temperature studies are still lacking.
The shift of the implanted $\mu^+$ (Fig.~\ref{fig:K}(b) may show the saturation, but these results are ambiguous,
since the shift can equally be fit to the CW dependence. The results from another implanted probe ($^{12}$B)
are also not sufficiently accurate to determine the low temperature behaviour \cite{boron12}.
The implanted probes present similar charged perturbations to the host, and the hyperfine couplings
imply a similar coupling to the host conduction bands. However, there are subtle differences between $\mu^+$ and Li$^+$.
First, Li$^+$ has a $1s$ core (absent for the $\mu^+$) that yields a repulsive interaction with the surrounding Pt atoms
that will cause an increased lattice distortion. On the other hand, the light $\mu^+$ exhibits a large zero point motion that may
also enhance the surrounding lattice relaxation.
Perhaps more significant is the difference in the distribution of the screening charge
induced by the pointlike charged defect.
The screening response of the Pt conduction electrons (Friedel oscillations) consists
of a balancing unit of charge in the vicinity of the defect (Friedel sum rule), but the significantly
stronger potential \cite{potl} for the $\mu^+$ indicates that its screening cloud is considerably
more compact than that of the Li$^+$. 

In comparison, the $^{195}$Pt NMR Knight shift is also negative but much larger
($-2.9$~\% vs. $-145$~ppm at room temperature). This is due to the large negative core-polarization hyperfine
coupling to the $d$ band \cite{vdk-brom00}. 
The orbital shift of Pt nuclei is also large, estimated to be 0.38\% \cite{clogston}.
$K(T)$ follows the bulk susceptibility over a wide range in temperature [e.g., see curve in Fig. \ref{fig:K}(a)].
Previous muon $K_\mu$ measurements in Pt claim the tracking of the $K_\mu$ and the bulk $\chi$ \cite{gygax1},
consistent with our findings. Because of the relatively weak coupling, the $^8$Li and $\mu^+$ shifts are less sensitive to any weak
temperature dependence of the orbital shift.
For the same reason, it may also be easier to interpret $K$ in terms of contributions
from the $d$ and $s$ band components of $\chi$.

The low temperature saturation of the spin susceptibility of nearly magnetic metals is not fully understood.
The relatively low saturation temperature indicates that it is connected with a correspondingly
small energy scale (much less than the Fermi energy, $E_F$).
It seems reasonable that an impurity could alter both the magnitude and temperature dependence
of this local magnetic response.
A similar situation is encountered in Pd, where the low temperature dip in $\chi$ is not reflected in the $^8$Li shift \cite{TerryPd}.
It appears that, at low temperature, the $^8$Li$^+$ in Pt may also be sensing a perturbed magnetic behaviour.
However, the defect response is surprisingly simple, appearing to restore the high temperature CW
behavior of the susceptibility all the way down to low temperatures, i.e. simply relieving the tendency to saturate.
This is a strong clue to the origin of the saturation of $\chi$ of Pt.

There is no obvious explanation for this effect of Li$^+$ on the local $\chi$ within any of the theories that attempt
to explain Pt's $\chi(T)$ \cite{bands, paramag, moriya-book, kim88},
with the possibe exception of the spin fluctuation model \cite{TerryPd}.
Here we propose an alternate explanation. In the absence of the Li, $\chi$ has contributions
from both the wide $s$ band and the narrow $d$ band. We assume the former is Pauli-like and $T$ independent
and the latter follows the CW law. At high temperatures, these contributions remain independent, but we suggest
a small interband coupling exists with a characteristic energy scale on the order of 200 K, such that below the
corresponding temperature, the bands are no longer independent but are locked together. 
Apparently the $T$ independent $s$ band dominates and $\chi$ becomes $T$ independent.
This interband coupling must be weak enough that it doesn't substantially alter the overall band structure that is
consistent with low temperature quantum oscillations \cite{dkc78}.
Saturation by this mechanism is reminiscent (perhaps fortuitously) of the Kondo screening
of a local magnetic impurity by a degenerate conduction band,
where far below the Kondo temperature, the CW-like $\chi$ saturates \cite{alloul75}.
In this context, consider the effect of the interstitial Li$^+$. We suggest that the combined effects of lattice relaxation
and electronic screening cause a local disruption in the interband coupling thereby restoring the $d$ band
CW behaviour down to much lower temperatures.

In this context, it is interesting to compare with the effects of other impurities in Pt.
Lithium-Platinum alloys and intermetallic compounds are known \cite{LiPt}. $\chi$ of Pt is rapidly
{\it suppressed} by alloying with Li (or other main group elements \cite{mainPt}), in contrast to our $K$.
In a rigid band picture, this is attributed to a doping effect that moves the Fermi level away from the narrow
$d$ band density of states peak.
In our measurements, the number of implanted $^8$Li in the sample
at any time is extremely small (less than $10^6$, i.e. the dilute limit),
so we do not expect any measurable change in the overall E$_F$.
However, $\chi$ may be altered locally.
As mentioned above, the Li$^+$ attracts a screening cloud of one unit of charge
to its vicinity (the neighbouring Pt) altering the local electronic structure and thus the magnetic response.
Such behaviour is well-established for {\it nonmagnetic} transition metal substituents in Pt.
Depending on the impurity valence, the net effect on the {\it bulk} $\chi$ can be either positive or negative \cite{IS1},and the
relative temperature dependence doesn't change markedly. However, Pt NMR
reveals this impurity effect on the Pt host is inhomogeneous \cite{WK,IS2}.
Satellite lines of the spin 1/2 $^{195}$Pt NMR show the
Knight shift is modified over several near neighbour shells around the impurity, presumably due to a variation of
the local density of states as it heals away from the impurity perturbation towards the bulk value.
In nearly all cases, however, the satellites have a Knight shift that is {\it smaller in magnitude} than the bulk shift.
Presuming the Pt hyperfine coupling is unaltered, this indicates
the local $\chi$ near the impurity is {\it reduced}, in contrast to what we find for the $^8$Li shift.
Only for Rh is the local $\chi$ enhanced at nearby Pt \cite{IS1,IS2}.
In a few of these dilute alloys, the impurity NMR has been carried out as well \cite{IKS}.
However, the temperature dependence of the local $\chi$ (either from the perspective of the
impurity or the nearby host) has not been studied in detail. The relatively large changes in the Pt shift
near the impurity suggest that the temperature dependence might well be modified. 
Theoretical studies show perturbative effects of nonmagnetic impurities are expected \cite{nonmagtheory}, but
no detailed explanation of the local Pt response has been advanced.
It would be interesting to extend these meaurements to see if our finding of the
recovery of the CW dependence is general.
It would also be interesting to study intrinsic point defects in Pt (vacancies, interstitials) in this regard.
Such defects, induced by irradiation, can yield superconductivity in closely related Pd \cite{stritzker}
which is at odds with an increased magnetic response.

For spin lattice relaxation from a single band, the simplest case of the Korringa law,
the Korringa product, $\mathcal{K} = (T_1TK^2) = \hbar\gamma_e^2/(4\pi k_B\gamma_n^2)\equiv S$,
is a constant independent of temperature \cite{korringa}. $\mathcal{K}$ can be enhanced substantially,
even in simple metals, by electronic correlations \cite{slichterbook,TerryPd,titberil}.
For $^8$Li in Pt, though $K$ is $T$ dependent, the enhancement is relatively modest ($\mathcal{K} \approx 2S$) at room temperature
increasing to $\approx 5S$ at low temperatures.
Stronger correlations in the form of spin fluctuations may be invoked to explain the $T$ dependent susceptibility
of exchange enhanced paramagnetic metals such as Pt and Pd.
As these effects become even stronger, on the approach to metallic ferromagnetism,
one finds the Korringa law itself is modified: the Korringa slope $1/T_1T$ acquires a temperature dependence
that generally increases with decreasing $T$ \cite{vdk-brom00,moriya-book}.
As for $^8$Li in Pd, Fig. \ref{fig:t1} shows no evidence of this latter effect.
The $T$ linear Korringa dependence of $1/T_1$ down to low temperature shows
no indication of a deviation related to the distinct $T$ dependence of the shift,
suggesting that, while spin fluctuations may be important to the static uniform
$\chi(T)$, they are not playing a substantial
role in the spin relaxation. Since both $1/T_1$ and $K$ are linear in $T$, $\mathcal{K}$ is proportional to $T^2$,
an unexpected (though simple) result.

Rather than a naive application of the single band Korringa law, however, we should consider
separately the contributions of different bands.
When several bands cross $E_F$, the Korringa slope $1/T_1T$ consists of 
approximately independent contributions (similar to the shifts) \cite{takiPd}.
In this case, the overall Korringa slope should still follow a
generalized Korringa law,
\begin{eqnarray}
\frac{S}{T_1T} = \sum_{\mbox{bands}} k K^2,
\label{genkor}
\end{eqnarray}
where the weights $k$ for each band, sometimes known as ``disenhancement'' factors,
may be either larger or smaller than one \cite{vdk-brom00}.
Note that, although different contributions to the shift may have opposite sign [Eq. (\ref{shiftcomps})],
contributions to $1/T_1$ are all additive.
We suspect this is the case for the shift in Pt where the $s$ contribution is likely positive, partially cancelling the negative $d$ band shift.
Eq. (\ref{genkor}) admits a particularly simple resolution to the problem of a simultaneous linear temperature dependence
of both $K$ and $1/T_1$.
Namely, if the weighting factor $k$ for the $d$ band is much less than for the $s$ band, then
$1/T_1$ is dominated by the $s$ band and remains linear in $T$ even though the shift from the $d$ band varies with $T$.
In other words, the $T$ dependent $d$ band shift is effectively decoupled from $1/T_1$.
This is also consistent with the magnitude of the Korringa slope for $^8$Li being comparable
to the noble metals \cite{TerryAu}, but contrasts to the case of $^{195}$Pt NMR,
where there are significant contributions from both the $s$ and $d$ bands as well as
the orbital relaxation \cite{vdk-brom00}.

For magnetically enhanced metals like Pt, the weighting $k$
can be small because the shift senses the enhanced uniform static susceptibility at infinite wavelength ($q=0$),
while the spin lattice relaxation is determined by a sum over the entire Brillouin zone of the low frequency dynamic susceptibility
that may be substantially less enhanced.
Another factor determining $k$ in transition metals has to do with orbital degeneracy of the $d$ band.
The $d$ electron core-polarization hyperfine coupling is approximately the same
for each $d$ orbital, so only fluctuations that are in phase between degenerate bands contribute
effectively to spin relaxation \cite{NarathWeaver}.
A similar consideration may apply in the case of the $s$-$d$ coupling to the interstitial $^8$Li.
Indeed, such an analysis of the $^{195}$Pt NMR does
predict $k$ for the $d$ band is less than 10\% that of the $s$ band.
Detailed calculations are required to confirm whether such a small value of $k$ for the
$d$ band is reasonable for Li.

\section{Conclusions}

To summarize, we measured the Knight shift and spin-lattice relaxation rate of dilute $^8$Li$^+$ implanted
in high purity Pt.  Below 300 K, the bulk susceptibility is approximately Pauli-like ($T$ independent), but exhibits a slight
temperature dependence and, below $20$~K, a minor magnetic impurity contribution, consistent with previous measurements.
In contrast, the Knight shift varies approximately linearly with temperature. It does not scale with $\chi$.
This discrepancy suggests the magnetic response of Pt is locally modified by the interstitial $^8$Li$^+$
similar to the case in Pd \cite{TerryPd}. Local perturbation of Pt's magnetic response is well established for nonmagnetic
transition metals. However, we have found that (at least for the case of $^8$Li) the perturbed response adopts
a particularly simple form, namely the extension of the high temperature Curie-Weiss behaviour down to low temperature.
Such a distinct isolated impurity response may shed light on the origin of the temperature dependent susceptibility of Pt and to transition metals in general.
From this, the single band Korringa product $\mathcal{K}$ that varies approximately as $T^2$ at low temperature.
This unusual result might be explained by a detailed theoretical calculation of the exchange enhanced metallic state.
On the other hand, the Korringa slope, consisting of contributions from both the $s$ and $d$ bands, may simply be dominated
by the $s$ band. In this case, one can obtain $1/T_1$ and $K$ which are both linear in temperature.

We acknowledge R. Liang at UBC and staff of TRIUMF for technical assistance, I. Elfimov and
B. Ramshaw for helpful discussions, and D.E. MacLaughlin for a critical reading of the manuscript.
This research is partially supported by NSERC of Canada and (through TRIUMF) by NRC of Canada and by CIFAR.


\begin{references}
\bibitem{chi-fit} The Curie-Weiss fit was done for the data at 250 K and above and is
\[
\chi_{CW} = \frac{0.00103(6)}{T+819.5(48.) K} + 5.32(1.9) \times10^{-8} \mbox{~cm$^3$/g}.
\]
The following form, including the low temperature saturation, fits the measured $\chi(T)$ well over the entire temperature range.
\[
\chi_p = a \frac{ [\theta/T^* + \tanh(T/T^*)]}{T+\theta} + \chi_0,
\]
where $a = 0.00023(1)$, $\theta = 569(76)$ K, $T^* = 240.3(16.1)$ K, and $\chi_0 =1.248 (.253) \times10^{-7}$~cm$^3$/g.
\bibitem{potl} e.g., The ionization energy of the corresponding neutral Li is much smaller than for muonium due to screening by
the core electrons.
\end{references}
\end{document}